%% file: 0_main.tex
\documentclass[lettersize,journal]{IEEEtran}
\usepackage{amsmath,amsfonts}
\usepackage{algorithmic}
\usepackage{array}
\usepackage[caption=false,font=normalsize,labelfont=sf,textfont=sf]{subfig}
\usepackage{textcomp}
\usepackage{stfloats}
\usepackage{url}
\usepackage{verbatim}
\usepackage{graphicx}
\def\BibTeX{{\rm B\kern-.05em{\sc i\kern-.025em b}\kern-.08em
    T\kern-.1667em\lower.7ex\hbox{E}\kern-.125emX}}
\usepackage{balance}
\usepackage{colortbl}
\usepackage{orcidlink}
\usepackage{kotex}
\usepackage{mathptmx}
\usepackage{booktabs}
\usepackage{multirow}
\usepackage{xcolor}
\graphicspath{{figs/}{figures/}{pictures/}{images/}{./}}

\begin{document}

\title{Discrepancies in Mental Workload Estimation: Self-Reported versus EEG-Based Measures in Data Visualization Evaluation}

\author{Soobin Yim\orcidlink{0000-0002-9887-8537}, Sangbong Yoo\orcidlink{0000-0002-0973-9288},~\IEEEmembership{Member,~IEEE}, Chanyoung Yoon\orcidlink{0000-0002-9784-0238}, Chanyoung Jung\orcidlink{0009-0001-2284-3506}, Chansoo Kim,\\Yun Jang\orcidlink{0000-0001-7745-1158},~\IEEEmembership{Member,~IEEE}, and Ghulam Jilani Quadri\orcidlink{0000-0002-8054-5048},~\IEEEmembership{Member,~IEEE}
\thanks{S. Yim, C. Yoon, and Y. Jang are with Sejong University, Seoul, South Korea. E-mail: \{tn12qls\,$|$\,vfgtr8746\}@gmail.com, jangy@sejong.edu}
\thanks{S. Yoo and C. Kim are with AI, Information and Reasoning (AI/R) Laboratory, Korea Institute of Science and Technology (KIST), Seoul, South Korea (e-mail: \{usangbong\,$|$\,eau\}@kist.re.kr)}
\thanks{C. Jung is with the Data Science Group, INTERX, Republic of Korea. Email: cy.jung@interxlab.com}
\thanks{G. J. Quadri is with the University of Oklahoma. Email: quadri@ou.edu}
\thanks{Y. Jang is the corresponding author.}}

\markboth{IEEE Transactions on Visualization and Computer Graphics,~Vol.~xx, No.~xx, xx~20xx}%
{Yim \MakeLowercase{\textit{et al.}}: Discrepancies in Mental Workload Estimation: Self-Reported versus EEG-Based Measures in Data Visualization Evaluation}

\maketitle

\begin{abstract}
    Accurate assessment of mental workload (MW) is crucial for understanding cognitive processes during visualization tasks. While EEG-based measures are emerging as promising alternatives to conventional assessment techniques, such as self-report measures, studies examining consistency across these different methodologies are limited. In a preliminary study, we observed indications of potential discrepancies between EEG-based and self-reported MW measures. Motivated by these preliminary observations, our study further explores the discrepancies between EEG-based and self-reported MW assessment methods through an experiment involving visualization tasks. In the experiment, we employ two benchmark tasks: the Visualization Literacy Assessment Test (VLAT) and a Spatial Visualization (SV) task. EEG signals are recorded from participants using a 32-channel system at a sampling rate of 128 Hz during the visualization tasks. For each participant, MW is estimated using an EEG-based model built on a Graph Attention Network (GAT) architecture, and these estimates are compared with conventional MW measures to examine potential discrepancies. Our findings reveal notable discrepancies between task difficulty and EEG-based MW estimates, as well as between EEG-based and self-reported MW measures across varying task difficulty levels.
    Additionally, the observed patterns suggest the presence of unconscious cognitive effort that may not be captured by self-report alone.
\end{abstract}



\input{1_intro.ins}
\input{2_relatedwork.ins}
\input{3_settings.ins}
\input{4_Preliminary_Study.ins}
\input{5_mainstudy.ins}
\input{6_discussion.ins}
\input{7_conclusion.ins}


\bibliographystyle{IEEEtran}

\bibliography{template}

\begin{IEEEbiography}[{\includegraphics[width=1in,height=1.25in,clip,keepaspectratio]{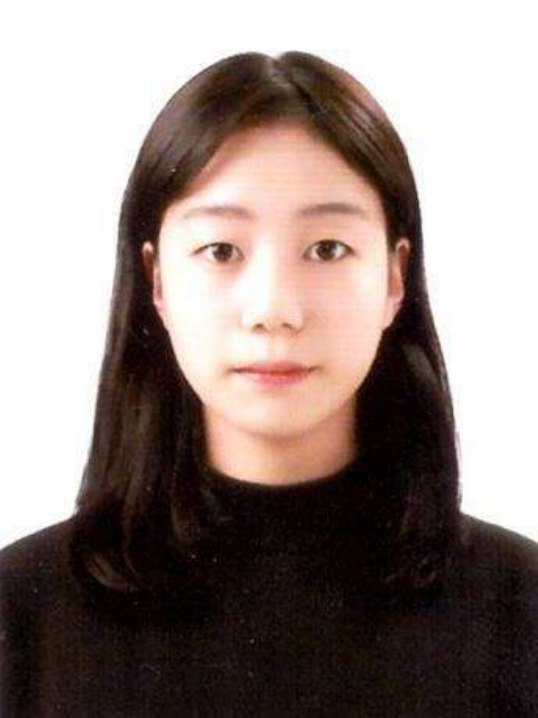}}]{Soobin Yim} 
    received the bachelor's and master's degrees in computer engineering from Sejong University, Seoul, Republic of Korea, in 2022 and 2024, respectively. She is currently a Ph.D. course student in convergence engineering for artificial intelligence and convergence engineering for intelligent drone at Sejong University, Seoul, Republic of Korea. Her research interests include visualization evaluation, EEG analysis, visual analytics, and traffic analysis.
\end{IEEEbiography}

\begin{IEEEbiography}[{\includegraphics[width=1in,height=1.25in,clip,keepaspectratio]{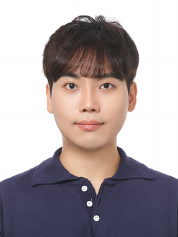}}]{Sangbong Yoo}
    received the bachelor's and Ph.D. degree in computer engineering from Sejong University, Seoul, South Korea, in 2015 and 2022, respectively. He was a postdoctoral researcher at Sejong University, from 2022 to 2025. He is currently a postdoctoral researcher at Korea Institute of Science and Technology (KIST), Seoul, South Korea. His research interests include information visualization, visual analytics, eye-gaze analysis, data quality, and volume rendering.
\end{IEEEbiography}

\begin{IEEEbiography}[{\includegraphics[width=1in,height=1.25in,clip,keepaspectratio]{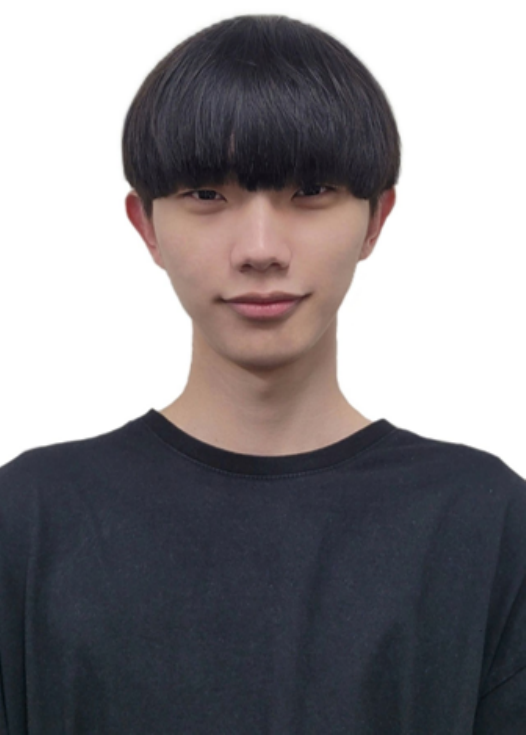}}]{Chanyoung Yoon}
    received a bachelor's and master's degree in computer engineering from Sejong University, Seoul, South Korea, in 2022 and 2023, respectively. He is now pursuing a Ph.D. degree at Sejong University, focusing on reinforcement learning and visual analytics. His research aims to deep reinforcement learning, visual analytics, and traffic prediction.
\end{IEEEbiography}

\begin{IEEEbiography}[{\includegraphics[width=1in,height=1.25in,clip,keepaspectratio]{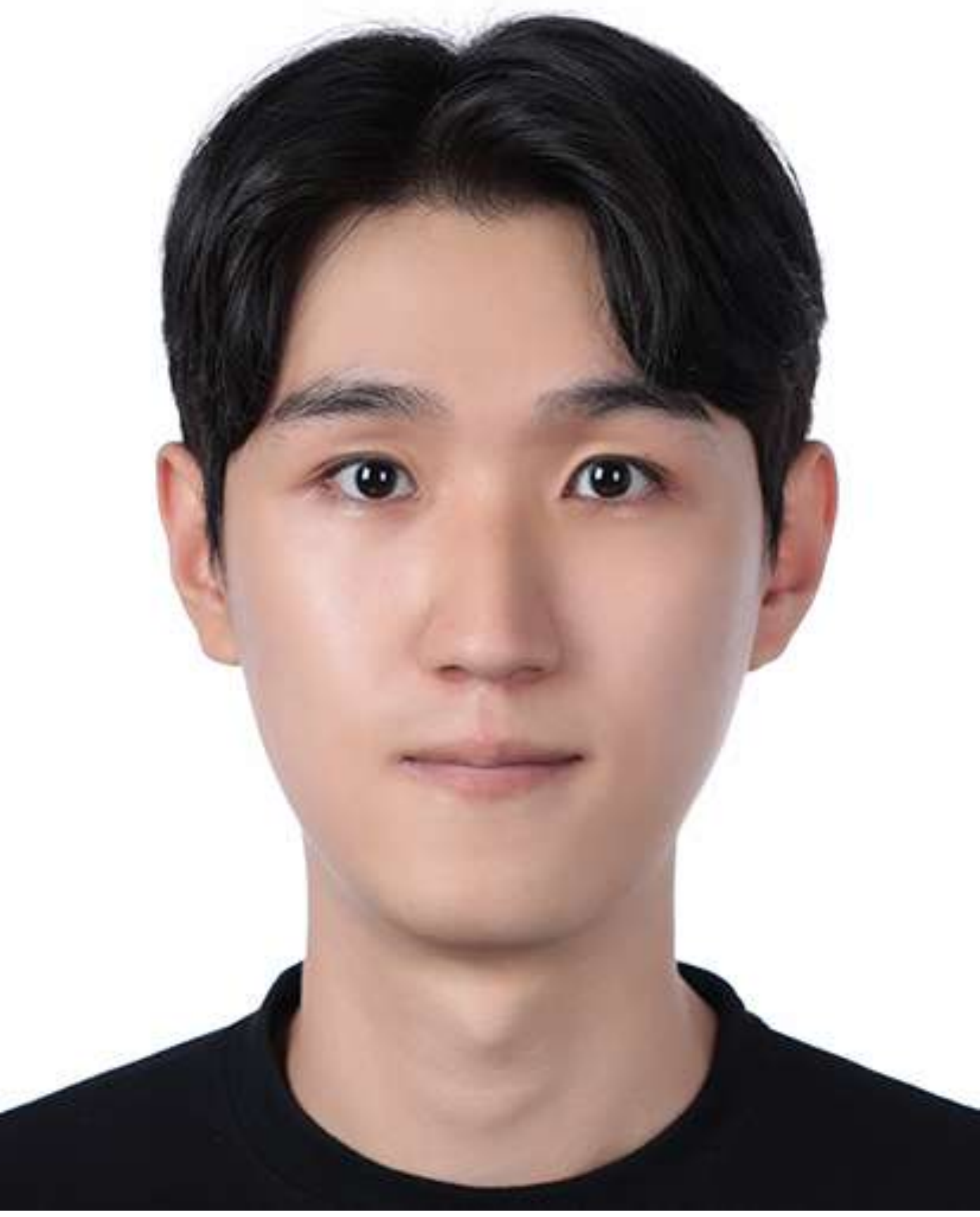}}]{Chanyoung Jung} 
    earned a BS in mathematics in 2022 and an MS in computer engineering and convergence engineering for intelligent drones, both at Sejong University, South Korea. Since 2024, he has been working as a researcher in a data science group called INTERX, Seoul, South Korea. His research interests include machine learning, predictive modeling and visual analytics.
\end{IEEEbiography}

\begin{IEEEbiography}[{\includegraphics[width=1in,height=1.25in,clip,keepaspectratio]{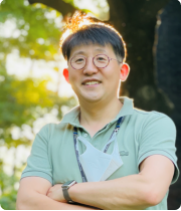}}]{Chansoo Kim}
    is an assistant professor at the University of Science and Technology and a senior research scientist at the Computational Science Centre, Korea Institute of Science and Technology. He leads the AI/R (AI, Information \& Reasoning) Lab., which explores the theoretical foundations of AI, the science of information, and complex (adaptive) systems. His researches span ethics and alignment, optimization, decentralization, and causality in AI—ranging from mathematical theory to real-world applications. Prof. Kim’s work centers on non-Gaussian behaviors—particularly heavy-tailed and leptokurtic—and their applications in learning, inference, finance, and inequality. He traces his academic lineage to C. F. Gau{\ss}. While grounded in theoretical AI, his lab’s research has also informed public policy. During the COVID-19 pandemic, the group supported the Korean CDC and the Office of the President with AI-driven, large-scale agent-based modeling.
\end{IEEEbiography}

\begin{IEEEbiography}[{\includegraphics[width=1in,height=1.25in,clip,keepaspectratio]{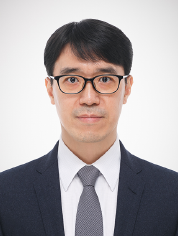}}]{Yun Jang}
    received the bachelor’s degree in electrical engineering from Seoul National University, South Korea, in 2000, and the master’s and Ph.D. degrees in electrical and computer engineering from Purdue University, in 2002 and 2007, respectively. He was a Postdoctoral Researcher at CSCS and ETH Z\"{u}rich, Switzerland, from 2007 to 2011. He is currently a professor in computer engineering at Sejong University, Seoul, South Korea. His research interests include interactive visualization, volume rendering, and visual analytics.
\end{IEEEbiography}

\begin{IEEEbiography}[{\includegraphics[width=1in,height=1.25in,clip,keepaspectratio]{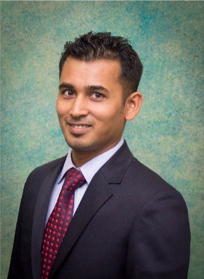}}]{Ghulam Jilani Quadri} is an Assistant Professor in the School of Computer Science at the University of Oklahoma. Prior to that, Dr. Quadri was a CIFellow postdoc at the University of North Carolina at Chapel Hill. He received his Ph.D. from the University of South Florida. Quadri received the 2021 Computing Innovation Fellow award. His research interests include creating human-centered frameworks to optimize visualization design and improve decision-making quality.
\end{IEEEbiography}
\end{document}

%% file: 1_intro.ins
\section{Introduction}
\begin{figure*}[tb]
    \centering
    \includegraphics[width=\linewidth]{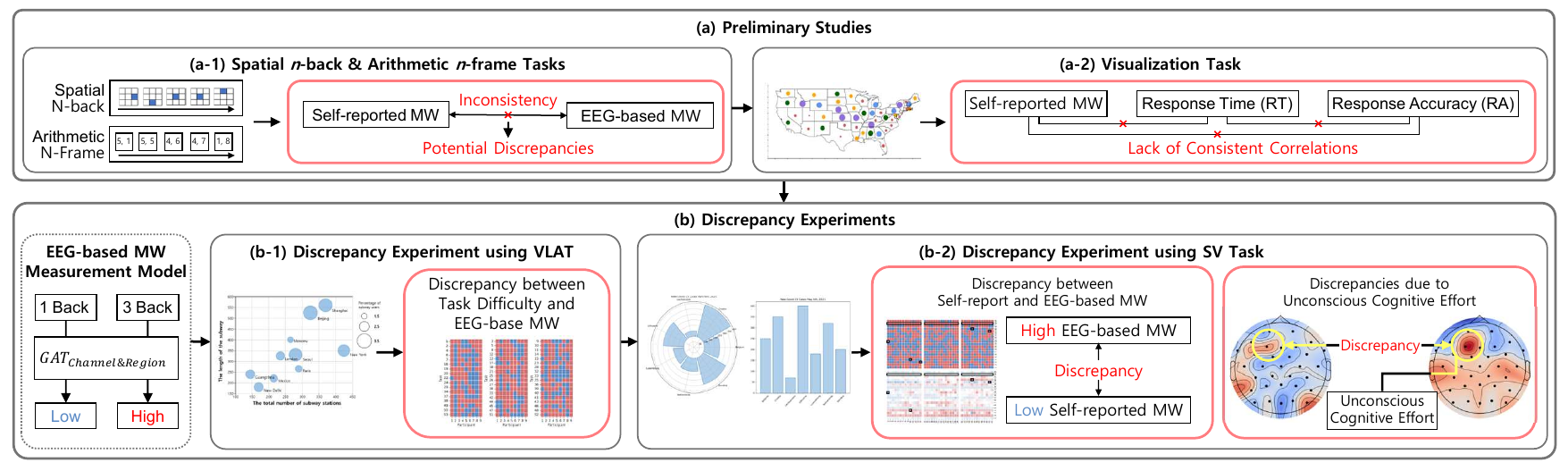}
    \caption{
        Our study sequentially investigates the consistency between self-reported and EEG-based MW. First, we examine inconsistencies among conventional MW measures in (a), followed by an analysis of discrepancies between self-reported and EEG-based MW during visualization tasks in (b). (a) presents preliminary studies examining the consistency of conventional MW measures. Spatial $n$-back and Arithmetic $n$-frame tasks are used in (a-1), and a visualization task is used in (a-2). (b) presents experiments designed to investigate discrepancies in MW estimation within the data visualization. The Visualization Literacy Assessment Test (VLAT) task is used in (b-1), and the Spatial Visualization (SV) task in (b-2).
    }
    \label{fig:overview}
\end{figure*}

\IEEEPARstart{M}{ental} workload (MW) refers to the portion of cognitive capacity allocated to task performance. Elevated MW during tasks demanding sustained attention has been shown to increase the likelihood of critical performance errors~\cite{dimitrakopoulos2017task}. Consequently, accurate assessment of MW is critical for understanding and optimizing human performance in complex tasks~\cite{byrne2011measurement}. Conventional assessments of MW often rely on subjective self-report tools, such as the NASA Task Load Index (NASA-TLX)~\cite{hart1988development}. EEG-based estimation~\cite{yim2023mental, yoghourdjian2020scalability}, increasingly used in conjunction with self-reported measures~\cite{castro2021examining, foo2013evaluating} for data visualization evaluation, has emerged as an alternative or complementary approach.

Nonetheless, empirical evidence supporting its consistency with self-reported assessments remains limited. Several studies have reported conflicting findings regarding the correspondence between EEG-based and self-reported MW measures. For instance, Mondellini et al.~\cite{mondellini2024multimodal} found that the two measures may diverge in virtual reality contexts, depending on the measurement devices used. In contrast, Lim et al.~\cite{lim2025neural} observed consistent trends between EEG-based and self-reported MW during robotic surgery tasks. Given the variability of findings across experimental contexts, it is indispensable to examine whether discrepancies between conventional and EEG-based measures of MW also emerge in visualization tasks.

In this paper, we investigate discrepancies between self-reported and EEG-based MW in the visualization domain. As illustrated in Figure~\ref{fig:overview}, our study consists of preliminary studies in (a) and discrepancy experiments in (b). Preliminary study (a-1) indicates potential discrepancies between EEG-based and self-reported MW measures. Preliminary study (a-2) reveals a lack of consistent correlations among conventional MW measures, including response time (RT), response accuracy (RA), and self-reported MW. Based on these findings, we design the discrepancy experiment quantitatively to investigate discrepancies between EEG-based and self-reported MW through two benchmark tasks: the Visualization Literacy Assessment Test (VLAT)~\cite{lee2016vlat} in (b-1) and the Spatial Visualization (SV) Task~\cite{tandon2022measuring} in (b-2). The VLAT task compares EEG-based MW with task difficulty, while the SV task compares EEG-based and self-reported MW.

We employ a personalized GAT-based EEG model for each participant. The model estimates MW by learning regional features and channel structures from EEG signals, which are represented as graphs.
The experiments reveal discrepancies involving EEG-based MW and task difficulty, self-reports and EEG-based MW, and unconscious cognitive effort. The main contributions of this paper are as follows.
\begin{itemize}
    \item We conduct empirical studies to examine the discrepancies between EEG-based and self-reported MW measures.
    \item We categorize discrepancies in MW with respect to task difficulty, self-reported assessments, and unconscious cognitive effort within the context of visualization tasks.
    \item We propose a personalized Graph Attention Network (GAT) model for MW estimation that learns both regional and channel-level features from EEG signals.
\end{itemize}

%% file: 2_relatedwork.ins
\section{Related Work}
\label{sec:relatedwork} 
In data visualization evaluation studies, MW measurement serves as a critical metric for quantifying the effectiveness of visualization techniques and assessing user experience~\cite{foo2013evaluating}. MW data is collected through various approaches, including interview-based subjective evaluations~\cite{kim2024phenoflow}, behavioral measurements such as response time (RT) and response accuracy (RA)~\cite{howard2020multi}, self-reported questionnaires~\cite{castro2021examining, foo2013evaluating, pollack2020association}, and physiological signal acquisition~\cite{yim2023mental, anderson2011user, nuamah2020evaluating}. One of the most widely used methods for assessing MW is the NASA Task Load Index (NASA-TLX)~\cite{hart1988development}, which has been extensively applied in the evaluation of visualization techniques~\cite{castro2021examining, pollack2020association, brown2019usability} and the comparison of visualization designs~\cite{yoghourdjian2020scalability, foo2013evaluating}. Recently, EEG-based biosignals have emerged as a promising tool for measuring MW, as EEG provides neural signals closely associated with cognitive processing, enabling a more direct and quantitative assessment of cognitive load~\cite{yoghourdjian2020scalability, anderson2011user, nuamah2020evaluating}. In particular, the application of deep learning techniques to EEG-based MW estimation is expanding~\cite{qiao2020ternary, kwak2020multilevel, guan2022eeg}, and such EEG-based models are newly being adopted in the evaluation of data visualizations~\cite{yim2023mental}.

In visualization evaluation, the adoption of EEG for MW measurement lacks sufficient reliability validation, prompting researchers to validate EEG-based MW approaches through multiple analytical frameworks, including correlation analysis with visualization task difficulty~\cite{yoghourdjian2020scalability, anderson2011user}, comparative assessment of evaluation patterns against NASA-TLX~\cite{nuamah2020evaluating}, and model performance benchmarking~\cite{yim2023mental}.
However, current EEG-based MW validation methods employed in visualization evaluation present several limitations. First, validation approaches utilizing correlation with visualization task difficulty lack reliability in their verification process, as task difficulty levels are arbitrarily determined by researchers rather than being established through extensive prior research with proven reliability~\cite{yoghourdjian2020scalability, anderson2011user}. Second, while validation methods comparing evaluation trends with NASA-TLX are reasonable, given that NASA-TLX represents a widely used self-reported survey for MW measurement, these approaches are limited by their reliance solely on qualitative comparisons without statistical significance testing~\cite{nuamah2020evaluating}. Third, model performance-based validation methods employ NASA-TLX as ground truth labels for EEG-based MW classification models, indirectly assessing the validity of EEG-based MW estimation through classification performance. However, since these EEG-based MW classification models are trained using subjective NASA-TLX ratings as reference standards, the objectivity of MW measurement remains constrained~\cite{yim2023mental}.
NASA-TLX represents a subjective assessment based on participants' post-task recollection. In contrast, EEG involves the collection of physiological signals in real time during task execution, which may result in discrepancies between these two metrics. Therefore, researchers validate EEG-based MW by statistically comparing whether MW measured by NASA-TLX and EEG-based MW demonstrate significant concordance.

Empirical studies examining the relationship between NASA-TLX scores and EEG-based MW have produced inconsistent and often contradictory results.
Studies reporting correlations between NASA-TLX scores and EEG-based MW have identified that only specific channels among all EEG channels exhibit correlations, with variations in correlated channels and frequencies depending on task difficulty~\cite{devos2020psychometric, aksu2024mental, zahabi2025cognitive}. 
Aksu et al.~\cite{aksu2024mental} discovered that theta power from AF4 and F8 channels, along with the $\frac{\text{theta}}{\text{alpha}}$ power ratio from AF3, AF4, and F8 channels, demonstrated strong correlations with NASA-TLX scores. Zahabi et al.~\cite{zahabi2025cognitive} compared correlations between NASA-TLX subscales (Mental Demand, Physical Demand, Temporal Demand, Performance, Effort, and Frustration) and EEG frequency bands. They found that alpha power correlated with NASA-TLX physical demand, while theta power showed significant correlations with NASA-TLX mental demand, performance, and frustration. Devos et al.~\cite{devos2020psychometric} identified that EEG channels exhibiting correlations between NASA-TLX scores and Event-Related Potential (ERP) P300 varied according to task difficulty.
Conversely, studies reporting no correlation between NASA-TLX scores and EEG-based MW suggest that these two metrics may reflect different dimensions of mental workload. 
Mondellini et al.~\cite{mondellini2024multimodal} found that NASA-TLX scores and EEG-based MW showed no significant correlation and responded differently under varying experimental conditions. Specifically, they observed that task difficulty led to significant MW differences only in the NASA-TLX scores, whereas unconscious MW differences related to modality were captured solely by EEG-based measures.

The presence and degree of correlation between EEG-based MW and NASA-TLX scores vary across studies, indicating a potential discrepancy between the two measures. 
However, prior studies on EEG-based MW estimation in visualization contexts frequently use NASA-TLX as a validation benchmark, often without explicitly addressing the potential discrepancies between subjective and EEG-based measures.
Therefore, to establish EEG-based MW as a reliable metric for assessing mental workload in visualization tasks, it is essential to examine whether such a discrepancy exists between EEG-based and self-reported measures. In this study, we investigate whether a discrepancy exists between EEG-based MW and self-reported MW in the context of visualization evaluation.

%% file: 3_settings.ins
\section{Experiment Setup for EEG Data}
\label{sec:dataRecording}
\begin{figure}[tb]
    \centering
    \includegraphics[width=\linewidth]{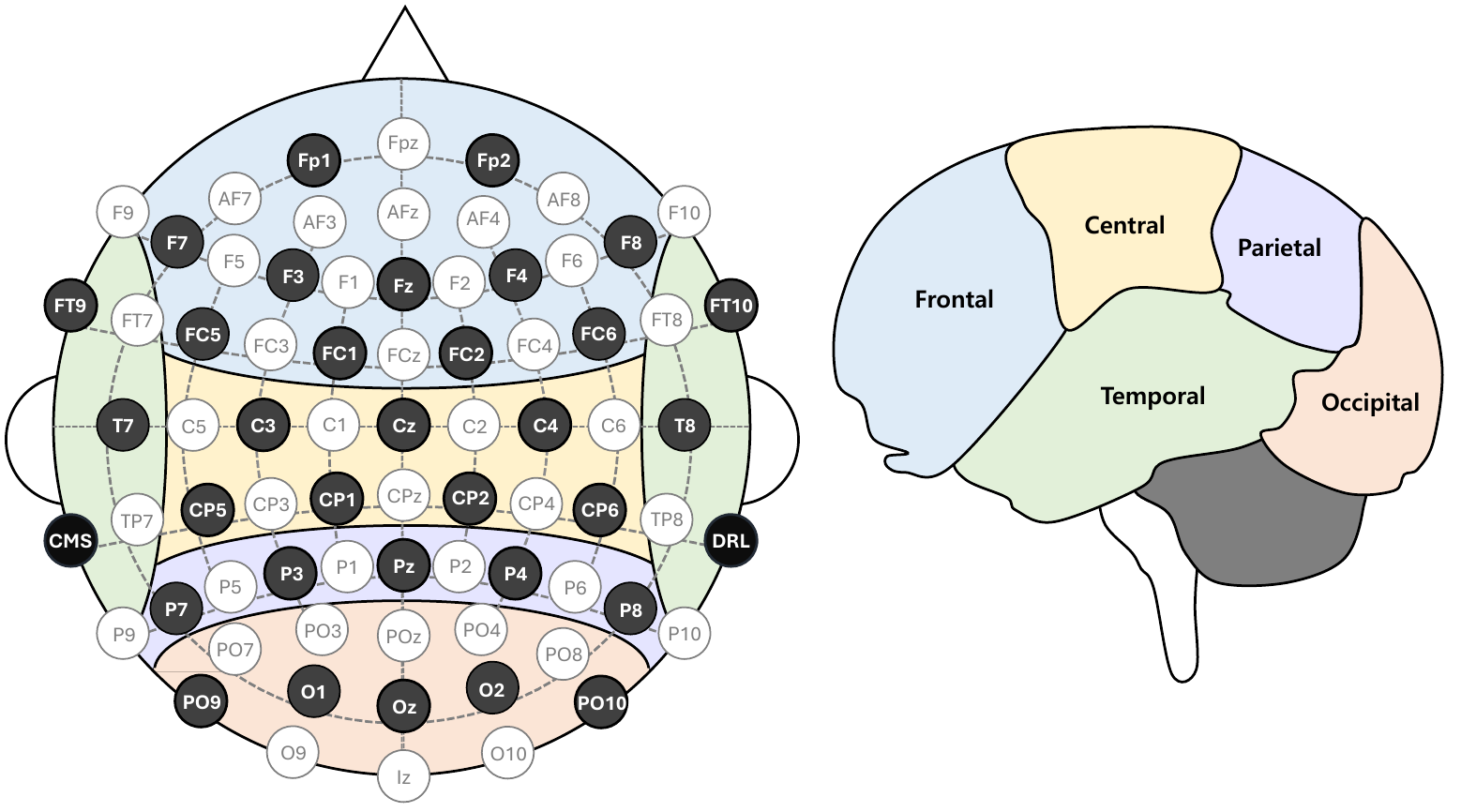}
    \caption{
        The Emotiv Epoc Flex sensor locations. All electrodes are mounted on the EasyCap, configurable to the international 10-20 system.
    }
    \label{fig:sensorPlacement}
\end{figure}

EEG data were collected using the Emotiv Epoc FLEX device equipped with saline-based sensors. The device consists of 32 Ag/AgCl electrodes (16 on each side), incorporating a common-mode sensor (CMS) on the left and a driven-right-leg (DRL) sensor on the right, which function as online references in active electrode EEG configurations~\cite{williams2020validation}. All electrodes were mounted on an EasyCap configured according to the international 10–20 system, using saline-soaked felt pads. Reference electrodes were placed on the earlobes, as illustrated in Figure~\ref{fig:sensorPlacement}.
The device records EEG data at a sampling rate of 128 Hz (internally 1,024 Hz), with a 14-bit resolution (1 least significant bit = 0.51$\mu$V), utilizing a 16-bit analog-to-digital converter, from which 2 bits corresponding to instrumental noise are discarded.

The built-in EEG signal processing pipeline includes a high-pass filter at 0.2 Hz, a low-pass filter at 45 Hz, digitization at 1,024 Hz, filtering with a digital 5th-order synchronization filter, and subsequent downsampling to 128 Hz. Prior to data acquisition, electrodes were prepared with 2–3 $mL$ of saline solution, and impedance was reduced below 20k~$\Omega$, as verified by the green status indicator in Emotiv Pro software (3.4.1).

EEG data were collected exclusively during task execution, excluding rest periods. The resulting dataset comprised 257 attributes, including timestamps, raw channel signals, contact quality (CQ), and signal quality (SQ).
Given the high sensitivity of EEG signals to participant movement, physical motion was minimized during data collection, limited to mouse clicks required for task responses. Answer buttons were fixed in position on the screen to reduce eye and hand movement. Participants viewed the stimuli on a 30-inch monitor ($1920 \times 1200$ resolution, 120 Hz refresh rate), positioned approximately 70 cm from the eyes.

For data preprocessing, artifacts resulting from participants’ body and eye movements were removed using independent component analysis (ICA).
Moreover, the theta band signals (4–7 Hz), which is known to be related to MW, was extracted from each EEG channel via frequency analysis and utilized in the MW experiments~\cite{smith2001monitoring, aksu2024mental, zahabi2025cognitive}.
Data points with low contact quality (CQ), indicative of poor electrode attachment, were excluded. Additionally, a weighting factor $w=(SQ/2)+0.5$, derived from signal quality (SQ), was applied to emphasize high-quality EEG signals~\cite{jeong2019cybersickness}.
To address distributional differences across EEG channels, min–max normalization was applied to scale the features to a [0, 1] range, using $X_{i} = (x_{i}-x_{min})/(x_{max}-x_{min})$.

%% file: 4_Preliminary_Study.ins
\section{Preliminary Studies}
This section investigates the consistency and validity of multiple MW measures, including response time (RT), response accuracy (RA), self-reported MW, and EEG-based MW, using two experiments, as illustrated in Figure~\ref{fig:overview}~(a).
Preliminary study (a-1) analyzes potential discrepancies between self-reported MW and EEG-based MW during mental tasks.
Preliminary study (a-2) explores the relationships among RT, RA, and NASA-TLX scores in the visualization tasks.

\subsection{Comparing Self-Report and Theta Power Measures of MW}
\label{sec:experiment1}
The NASA-TLX is a widely used self-reported measure for assessing MW, and EEG-based MW assessments are increasingly recognized as objective methods that complement or replace self-reports. To control MW while minimizing the influence of visual encoding, this experiment employs Spatial $n$-back and Arithmetic $n$-frame tasks, as illustrated in Figure~\ref{fig:overview} (a-1). This setup enables the examination of inherent differences between self-reported MW and EEG-based MW, as well as the exploration of potential discrepancies between the two measures.

\subsubsection{Participants}
\label{sec:n-backExperimentMethod}
Four male participants aged 24 to 35 were recruited. All had at least a high school education; 
Two have a high school diploma, one holds a bachelor's degree, and another a doctorate. 
To ensure a consistent mental state, participants were instructed to sleep at least 8 hours and avoid alcohol for 48 hours before the experiment. The experiment was fully explained in person, and written informed consent was obtained from all participants.

\subsubsection{Tasks}
\label{sec:ps1_task}
In this experiment, we designed tasks based on the Spatial $n$-back and Arithmetic $n$-frame tasks utilized by Zhang et al.~\cite{zhang2018learning}.
The participants performed Spatial 1-back, 3-back, Arithmetic 1-frame, and Arithmetic 2-frame tasks. Each task consisted of 200 stages, each taking 3 seconds to complete. Therefore, the participants spent 10 minutes on each task. Additionally, we provided participants with a mandatory 5-minute break between tasks to minimize the impact of mental fatigue during task completion. Therefore, our experiment took 55 minutes per participant, including break time. Moreover, we provided sufficient pre-training to enhance the participants' understanding of each task. After completing each of the four tasks, the participants responded to NASA-TLX surveys. EEG signals were recorded for 10 minutes while participants performed tasks.

\textbf{Spatial $n$-Back Task: }
The Spatial $n$-back task requires the participant to observe a grid of nine square blocks and recall the location of the blue block that appeared $n$ steps earlier. As $n$ increases in the Spatial $n$-back task, participants need to remember more location sequences, increasing their MW. Zhang et al.~\cite{zhang2018learning} explained that they could discern the MW induced by only spatial 1-back and 3-back tasks. Therefore, in our experiment, we used the 1-back and 3-back tasks. 
In the task, the blue block appears on the screen for 0.5 seconds at each step and then becomes inactive. 
The inactive state is maintained for 2.5 seconds before the blue block reappears at a new location, randomly assigned from among the nine white blocks. For the 1-back task, the participants were asked to tell if the current and previous locations of the blue block were the same. For the 3-back task, the participants were asked to tell if the current and three steps prior locations of the blue block were the same. The participants responded by pressing the correct button for the correct frame. We collected EEG data, RT, and RA during the task.

\textbf{Arithmetic $n$-Frame Task: }
The Arithmetic $n$-frame task is an experiment in which participants are presented with pairs of random numbers between 1 and 9 on the screen and are required to remember the sum of numbers in $n$ frames, where $n$ is the number of frames. 
Similar to the Spatial $n$-back task, increasing the value of $n$ imposes a greater mental workload on participants, as they must retain and recall a larger number of prior frames. The two numbers displayed on each frame are randomly selected. In the 1-frame task, the participants need to choose the frame with a pair of numbers that add up to 10. In the 2-frame task, the participants need to select the frame with two pairs of numbers, one from the current frame and one from the previous frame, whose sum is 20. Like the Spatial $n$-back task, the participants observe the activated number for 0.5 seconds and respond during the following 2.5 seconds when the number is no longer active. The participants responded by pressing the correct button for the correct frame. We also collected EEG data, RT, and RA during this task.

\subsubsection{Potential discrepancies}
\label{sec:ps2result}

\begin{table}[t]
\caption{
    EEG channels that exhibit agreement between theta power and NASA-TLX scores, quantified using Cohen's Kappa statistic ($k$) after discretizing both variables into tertiles (33\%/67\%).
    EEG channels with the highest $k$ values are identified for each mental task.
}
\label{tab:kappa}
\resizebox{\columnwidth}{!}{
\begin{tabular}{|c|c|r|}
\hline
Task  & Channel  & \multicolumn{1}{c|}{$k$} \\ \hline
Spatial 1-back & Cz, Fz, FT9, CP5, P8 & 0.636 \\ \hline
Spatial 3-back & \begin{tabular}[c]{@{}c@{}}Cz, Fz, Fp1, F7, C3, FC5, FT9, \\ T7, CP5, CP1, PO9, O1, Pz, Oz,\\ T8, C4, F4, F8, Fp2, roi\_left, roi\_right \end{tabular} & 1.0   \\ \hline
Arithmetic 1-frame    & P4  & 1.0   \\ \hline
Arithmetic 2-frame    & Oz & 1.0   \\ \hline
\end{tabular}}
\end{table}

We compute the Spearman rank correlation coefficient ($r$) between theta power and NASA-TLX scores for each task and EEG channel. The theta power refers to the average power spectral density (PSD) of the theta band in the EEG data.
The theta power is a scalar measure that reflects the energy level within the theta frequency band (4–7 Hz) and is widely utilized as a neural marker for EEG-based MW estimation~\cite{castro2020validating}. The analysis includes all individual channels from the Emotiv Epoc Flex as well as two regions of interest (RoIs)~\cite{castro2020validating}: \textit{roi\_right} (F4 and Fp2) and \textit{roi\_left} (F3 and Fp1). Among all channel-task combinations, only Spatial 3-back (Fp1, Fp2, and \textit{roi\_right}) and Arithmetic 2-frame (Oz) exhibit statistically significant monotonic relationships ($r = 1.00$, $p < 0.001$). This correlation analysis suggests that the association between theta power and self-reported MW is not consistently observed across tasks and regions, implying limited generalizability of EEG-based MW correlations without task- or region-specific consideration.

Table~\ref{tab:kappa} present Cohen's Kappa coefficient ($k$). Cohen's Kappa coefficient ($k$) indicating the agreement between theta power and NASA-TLX scores. Both values are discretized using tertile thresholds (33\% / 67\%) for comparison. We summarize channels that show relatively high $k$ values for each mental task, including the Spatial $n$-back and Arithmetic $n$-frame tasks. Note that $k \geq 0.61$ indicates substantial agreement and $k \geq 0.81$ indicates almost perfect agreement~\cite{landis1977measurement}. Table~\ref{tab:kappa} lists only the tasks and channels for which $k \geq 0.61$.

In the 1-back task, the Cz, Fz, FT9, CP5, and P8 channels show substantial agreement ($k=0.64$). In the 3-back task, nearly all channels including Fp1, Fp2, F7, C3, FC5, FT9, T7, CP5, CP1, PO9, O1, Pz, Oz, T8, C4, F4, F8, \textit{roi\_left}, and \textit{roi\_right} exhibit almost perfect agreement ($k=1.00$). In contrast, only P4 in the Arithmetic 1-frame task and Oz in the Arithmetic 2-frame task reach this level of agreement ($k=1.00$). This agreement analysis suggests that the degree of agreement between EEG-based and self-reported MW may vary across different task types. While the 3-back task exhibits near-perfect agreement across many channels, this consistency is not observed in other tasks such as the Arithmetic 1-frame and 2-frame tasks.

These potential discrepancies analysis shows that consistency between theta power and NASA-TLX scores appears only in specific task and channel conditions. The lack of alignment in most cases suggests potential discrepancies between the two measurement approaches.

\subsection{Correlation between Conventional Measurements}
\label{sec:preliminaryStudy1}
This study examines the correlations between conventional MW measures, including RT, RA, and self-reported scores, in visualization tasks, using the NASA-TLX for self-reported assessment.
The analysis is conducted across tasks involving scatter plots, bar charts, line charts, and map visualizations, which were designed based on the visualization guidelines~\cite{yim2023mental}.

\subsubsection{Participants}
\label{sec:ps1_experiment_method}
We recruited ten participants, five male and five female students, aged between 23 and 33. Among them, five participants were graduate students majoring in data visualization, and the remaining five were undergraduate students who had completed a data visualization course.

\subsubsection{Tasks}
In each experimental trial, participants were simultaneously presented with a visualization and a corresponding task. They used the visualization to complete the task and recorded their responses via keyboard or mouse input. RTs were collected during this process.
After each task, participants completed a NASA-TLX survey to assess their perceived MW, followed by an initial 3-second rest period. During this resting phase, participants were allowed to extend their break time if needed. 
Once the participant felt ready to proceed, the next task began. Participants completed a total of 72 visualization tasks, following an initial practice phase that consisted of six visualization tasks. For the main experiment, tasks were structured according to four visualization types, each having three possible tasks, three distinct design guides, and two design guide violation conditions. Tasks were presented in random order to avoid learning effects or fatigue.

\subsubsection{Correlation}
\label{sec:correlationAnalysis}
We anlayzed the Spearman rank correlation coefficients ($r$) among RA, RT, and NASA-TLX scores. The correlation between RA and RT was not significant ($r=-0.17, p=0.106$). NASA-TLX shows a weak negative correlation with RA ($r = -0.35, p<0.005$) and a weak positive correlation with RT ($r = 0.69, p<0.005$), suggesting that MW, as measured by NASA-TLX, does not consistently align with RT and RA.

These correlation results indicate that conventional MW measures, such as, RT, RA, and NASA-TLX scores, are not strongly correlated, suggesting that each metric may reflect different facets of MW in visualization tasks.

\begin{figure*}[tb]
    \centering
    \includegraphics[width=\linewidth]{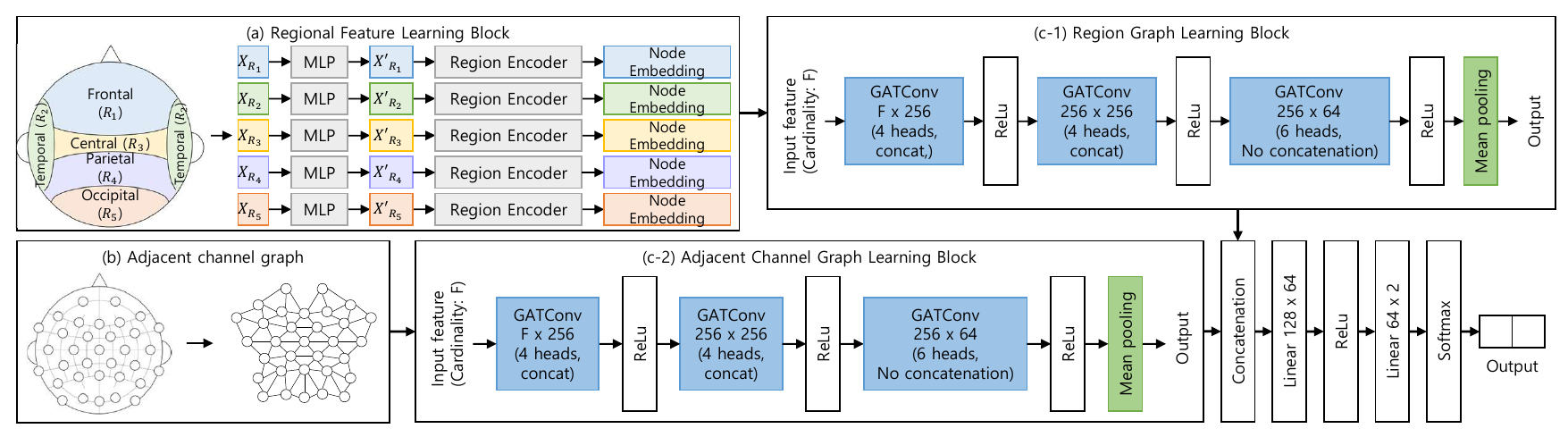}
    \caption{
        Architecture of the personalized Graph Attention Network (GAT) model. (a) is the regional feature learning block, while (b) is the adjacent channel graph. (c-1) and (c-2) depict the structures of the GAT, where (c-1) receives the graph from (a) and (c-2) takes the graph from (b) as input to estimate MW.
    }
    \label{fig:model}
\end{figure*}

\subsection{Findings and Discussion}
The preliminary studies assessed the potential discrepancy and consistency of multiple MW measurement approaches through correlation-based analyses. Preliminary study (a-1), detailed in Section~\ref{sec:ps2result}, examines the potential discrepancy between self-reported and EEG-based MW estimates. Preliminary study (a-2), described in Section~\ref{sec:correlationAnalysis}, investigates the relationships among conventional metrics, including RT, RA, and self-reported MW.

These analyses deliver two key insights. First, the consistency between EEG-based and self-reported MW measures appears limited and varies depending on the context. Second, the weak and inconsistent correlations among conventional MW indicators, such as RT, RA, and self-reports, suggest that these measures reflect different dimensions of MW rather than a unified construct. These results highlight the need for a multi-faceted approach to MW evaluation and suggest the presence of systematic discrepancies between subjective and physiological metrics. To further investigate these discrepancies, Section~\ref{sec:main_study} introduces a study involving visualization tasks of varying complexity and domains.

%% file: 5_mainstudy.ins
\section{Discrepancies in MW Measurements}
\label{sec:main_study}
This section presents two experiments designed to examine discrepancies between different MW measures, as illustrated in Figure~\ref{fig:overview} (b). Before describing the experiments, we introduce the EEG-based MW estimation model used to quantify MW during the experimental tasks. To investigate potential discrepancies in MW measurements, we conducted experiments using two task types: the Visualization Literacy Assessment Test (VLAT)~\cite{lee2016vlat} and the Spatial Visualization (SV) Task~\cite{tandon2022measuring}. The experiment based on VLAT, as illustrated in Figure~\ref{fig:overview} (b-1), examines the discrepancy between EEG-based MW and task difficulty. The SV Task-based experiment, shown in Figure~\ref{fig:overview} (b-2), explores two types of discrepancies: one between self-reported and EEG-based MW across varying task difficulties and another emerging from unconscious cognitive effort.

\subsection{EEG-based MW Measurement Model}
\label{sec:model_ex1}
EEG-based MW can be quantified through activation levels of specific cortical regions or EEG channels, as well as by analyzing functional connectivity patterns, including interactions and synchronization among multiple brain regions or channels. Therefore, to measure the interaction of each region or channel, we develop an architecture that learns adjacent channel and brain region features in Graph Attention Network (GAT), as shown in Figure~\ref{fig:model}.
Our model architecture comprises three main components: (a) a regional feature learning block~\cite{fan2024eeg} designed to capture both activation and interaction patterns within each brain region relative to MW; (b) an adjacent channel graph module that represents node-specific activations and interactions between neighboring EEG channels; and (c) two GAT learning blocks—one for the region graph (c-1) and another for the adjacent channel graph (c-2)—which process the respective graph structures. The outputs from the regional feature block (a) and the graph constructed in (b) are fed into the GAT models in (c-1) and (c-2), respectively. Finally, the model integrates these learned representations to evaluate MW.

In Figure~\ref{fig:model} (a), EEG channels are grouped into $n$ brain regions according to their anatomical sensor locations, as shown in Figure~\ref{fig:sensorPlacement}. Each region $R_i$ is defined as a set of channels, and its input is represented by $X_{R_i} = \{ x_{k_j} \mid k_j \in R_i \}$, where $X_{R_i} \in \mathbb{R}^{m \times T_s}$ and $x_{k_j}$ denotes the time-series signal of channel $k_j$ with time length $T_s$. This input is passed through a multilayer perceptron (MLP) to obtain a high-dimensional representation $X'_{R_i} = \text{MLP}(X_{R_i}) \in \mathbb{R}^{m \times F_{\text{in}}}$. A region encoder $f_i$ summarizes it into a single embedding vector $\mathbf{h}_i = f_i(X'_{R_i}) \in \mathbb{R}^{F}$. The resulting embeddings $\{\mathbf{h}_1, \mathbf{h}_2, \dots, \mathbf{h}_n\}$ serve as node features for the region-level graph in Figure~\ref{fig:model} (c-1). In Figure~\ref{fig:model} (b), a graph is constructed by connecting each node to its neighbors according to the 10-20 system. Let $V = \{v_1, v_2, \dots, v_N\}$ be the set of nodes; the edge set $E$ defines the graph $G = (V, E)$, used for training with the structure in (c-2).

Ground truth labels were assigned based on the Spatial $n$-back task, with 1-back trials representing low MW labeled as 0, and 3-back trials representing high MW labeled as 1. Each task is executed for the same amount of time, ensuring a balanced class distribution. A sliding window was employed to augment training data and facilitate stable model learning. 
The theta band signals, collected from 32 channels, were formatted into structured $32 \times 32$ matrices to serve as input for the model. We use theta band signals instead of theta power to observe inter-channel interactions while preserving phase information. To account for individual differences in cognitive processing, personalized MW estimation models were trained separately for each participant. Each model was trained using EEG data collected during the participant’s Spatial $n$-back task. According to Table~\ref{tab:kappa}, EEG data collected from the Arithmetic $n$-frame task is not used for model training due to inconsistencies in channel activation patterns across different MW levels.
The dataset was partitioned into 80\% for training and 20\% for testing, and model performance was assessed using 10-fold cross-validation to ensure robustness and generalizability.

\begin{table}[t]
\caption{
    Average Classification Performance of Personalized EEG-Based MW estimate Models
}
\label{tab:model_result_compare}
\resizebox{\columnwidth}{!}{
\begin{tabular}{|c|r|r|r|r|}
\hline
Metric      & \multicolumn{1}{c|}{SVM}           & \multicolumn{1}{c|}{$GAT_{Channel}$}    & \multicolumn{1}{c|}{$GAT_{Region}$}  & \multicolumn{1}{c|}{$GAT_{Channel\&Region}$}          \\ \hline
Accuracy (\%)    & 78.37 & 98.46 & 99.05 & 99.38 \\ \hline
Loss        & 0.4945   & 0.0574  & 0.0489  & 0.0397  \\ \hline
Sensitivity & 0.7428   & 0.9883  & 0.9919  & 0.9923  \\ \hline
Specificity & 0.8026   & 0.9807  & 0.9892  & 0.9952  \\ \hline
F1-Score    & 0.7827   & 0.9839  & 0.9890  & 0.9935  \\ \hline
\end{tabular}}
\end{table}

Table~\ref{tab:model_result_compare} presents the MW estimation performance of each model architecture, evaluated across all participants recruited as described in Section~\ref{sec:experiment2} and \ref{sec:experiment3}. Accuracy, loss, sensitivity, specificity, and F1-score are reported as average values across participants. We compare four models: a baseline SVM classifier using theta power, a conventional statistical approach; $GAT_{\text{Channel}}$ model, which learns channel features as shown in Figure~\ref{fig:model} (b) and (c-2); $GAT_{\text{Region}}$ model, which learns region features as shown in Figure~\ref{fig:model} (a) and (c-1); and $GAT_{\text{Channel\&Region}}$ model, which jointly learns both region and channel features. Among these, $GAT_{\text{Channel\&Region}}$ achieves the most robust MW estimation performance with the lowest standard deviation across participants. The $GAT_{\text{Channel}}$ is constrained to learning local correlations due to its static graph structure, while $GAT_{\text{Region}}$ is unable to capture intra-regional interactions as it merges all regional channels into a single node. In contrast, $GAT_{\text{Channel\&Region}}$ integrates both global and regional activation patterns, resulting in improved performance and interpretability. Based on these findings, we selected $GAT_{\text{Channel\&Region}}$ for discrepancy analysis experiments.

\begin{figure}
    \centering
    \includegraphics[width=\columnwidth]{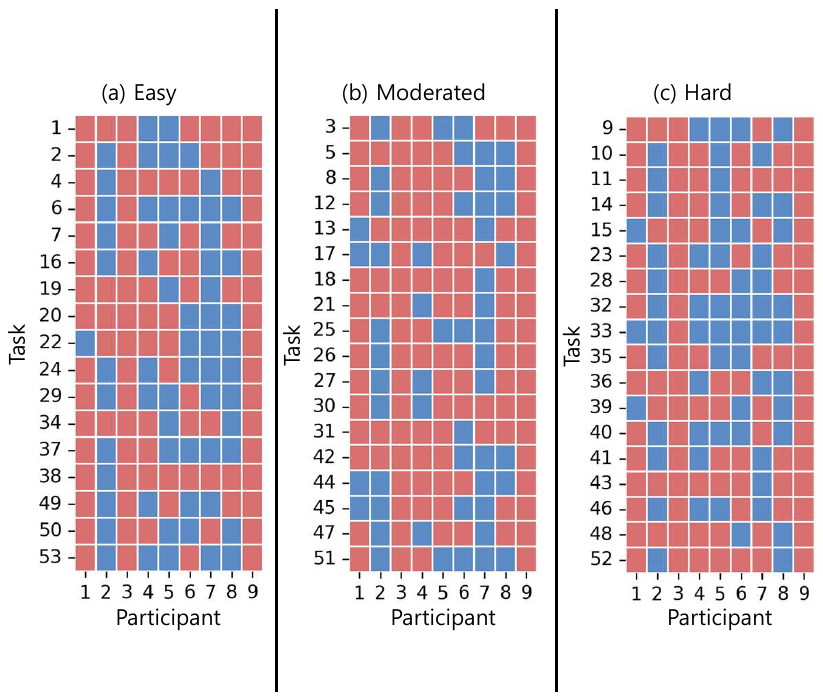}
    \caption{
        EEG-based MW estimation across task difficulty levels in the VLAT:
        (a) shows an easy task, (b) a moderate task, and (c) a hard task.
        EEG-based MW estimates remain unchanging across task difficulty levels, indicating a discrepancy from self-reported measures.
    }
    \label{fig:vlat}
\end{figure}

\subsection{Discrepancy Experiment using VLAT}
\label{sec:experiment2}
As illustrated in Figure~\ref{fig:overview} (b-1), we investigate discrepancies between EEG-based MW estimates and task difficulty using a benchmark derived from the VLAT~\cite{lee2016vlat}. 
VLAT was originally designed to evaluate individuals’ ability to interpret and reason about information presented in various visualization types and task formats commonly used in the field of information visualization.
This experiment aims to assess whether EEG-based MW measures consistently align with task difficulty across diverse visualization contexts.

\subsubsection{Participants}
Eight participants, five males and three females, aged between 24 and 35 years, were recruited. 
All participants have at least a high school education: four have a high school diploma, three hold a bachelor's degree, and one holds a doctorate.

\subsubsection{Tasks}
We used 53 test items from VLAT, 35 four-option multiple-choice items, three three-option multiple-choice items, and 15 true-false items. Participants received detailed instructions on the purpose and procedures of the experiment. No interactive techniques were included in the visualizations to prevent potential learning effects when encountering unfamiliar visualization interactions~\cite{boy2015suggested}, and to avoid motion-induced noise in EEG data. Participants were also instructed to skip any questions they did not know the answer to. EEG was collected while participants performed the tasks, using the method described in Section~\ref{sec:dataRecording}.

\begin{table}[t]
\caption{
    Comparison of Differences Between EEG Clusters using UMAP According to Task Difficulty Levels in the VLAT
}
\label{tab:cluster_vlat}
\resizebox{\columnwidth}{!}{
\begin{tabular}{|c|r|r|r|}
\hline
Metric                & \multicolumn{1}{c|}{(1) Easy-Moderated} & \multicolumn{1}{c|}{(2) Easy-Hard} & \multicolumn{1}{c|}{(3) Moderated-Hard} \\ \hline
Cohen's $d$              & 0.132             & 0.111        & 0.022             \\ \hline
Mahalanobis Distance   & 0.138             & 0.116        & 0.021             \\ \hline
Bhattacharyya Distance & 0.092             & 0.084        & 0.001             \\ \hline
Overlap Coefficient    & 0.611             & 0.598        & 0.743             \\ \hline
\end{tabular}}
\end{table}

\subsubsection{Discrepancy between EEG-based MW and Task Difficulty}
\label{sec:ex2result}
We analyze the discrepancies between EEG-based MW and task difficulty from two perspectives: the $GAT_{\text{Channel\&Region}}$ model and the EEG signals themselves, by comparing them against task difficulty.

Figure~\ref{fig:vlat} shows how the EEG-based MW estimated by the $GAT_{\text{Channel\&Region}}$ model correspond to task difficulty. It presents the EEG-based MW estimated for each participant during the task, with red indicating high MW and blue indicating low MW. Note that the $GAT_{\text{Channel\&Region}}$ model achieved high reliability across the eight participants, with an average accuracy of $99.09\%$, loss of $0.37$, sensitivity of $0.988$, specificity of $0.994$, and F1-score of $0.990$. As shown in Figure~\ref{fig:vlat}, the VLAT categorizes task difficulty into three levels: (a) easy, (b) moderate, and (c) hard. However, paired t-tests conducted on EEG-based MW estimations revealed no statistically significant differences across these difficulty levels, with all pairwise comparisons—(a) vs. (b), (a) vs. (c), and (b) vs. (c)—yielding $p$-values greater than 0.05.

From the perspective of EEG signals, we analyze the discrepancy by examining the distances between EEG data distributions across tasks with varying difficulty levels, as shown in Table~\ref{tab:cluster_vlat}.
This table presents how EEG signal distributions shift according to task difficulty in the VLAT.
The EEG signals are clustered using Uniform Manifold Approximation and Projection (UMAP)~\cite{mcinnes2018umap}.
For this analysis, we adopt four statistical measures: Cohen's $d$, which standardizes mean differences between two groups by pooled standard deviation; Mahalanobis distance, which measures the distance between distribution centers while accounting for data covariance; Bhattacharyya distance, which quantifies the similarity between two probability distributions; and the overlap coefficient, which estimates the actual overlapping area between two probability density curves based on kernel density estimation. Note that a Cohen's $d$ of 0.2 or greater is considered a small effect, while values below 0.2 indicate negligible differences~\cite{cohen2013statistical}. A larger Mahalanobis distance suggests greater statistical separability~\cite{mahalanobis2018generalized}, whereas a smaller Bhattacharyya distance indicates stronger overlap between distributions~\cite{kailath2003divergence}. The overlap coefficient approaches 1 when two distributions are nearly identical~\cite{inman1989overlapping}.
Table~\ref{tab:cluster_vlat} presents pairwise comparisons of EEG signal distributions across task difficulty levels: (1) easy vs. moderate, (2) easy vs. hard, and (3) moderate vs. hard. In all cases, Cohen's $d$ values indicate negligible effect sizes \big((1) = 0.132, (2) = 0.111, (3) = 0.022\big), and the corresponding Mahalanobis distances are similarly small \big((1) = 0.138, (2) = 0.116, (3) = 0.021\big), suggesting minimal separability between groups. Bhattacharyya distances also reveal substantial overlap in the distributions, especially in comparison (3), where the overlap is nearly complete. This pattern is further supported by the overlap coefficients \big((1) = 0.611, (2) = 0.598, (3) = 0.743\big), indicating a high degree of similarity in EEG signal distributions across different task difficulty levels.

These analyses indicate that EEG signal distributions show limited sensitivity to variations in task difficulty. This observation reveals a discrepancy between EEG-based MW estimation and task difficulty~\cite{lee2016vlat, hedayati2024university}.

\subsection{Discrepancy Experiment using Spatial Visualization Task}
\label{sec:experiment3}
As illustrated in Figure~\ref{fig:overview} (b-2), we examine two types of discrepancies in MW measurement: (1) the mismatch between task difficulty and EEG-based MW, and (2) the inconsistency between EEG-based and self-reported MW estimates. This experiment uses the SV Task as a benchmark, which involves visualizations with various orientations and rotations to assess participants' spatial reasoning abilities. We selected the SV Task because its diverse visual stimuli offer a wide range of complexity levels, making it well-suited for evaluating MW across different visual and cognitive demands.

\subsubsection{Participants}
\label{sec:vis_ex2_method}
Fourteen males and five females aged between 24 and 35 were recruited. 
All participants have at least a high school education: eleven have a high school diploma, seven hold a bachelor's degree, and one holds a doctorate.

\subsubsection{Tasks}
\label{sec:sv_task}
Each participant performed 60 Spatial $n$-back tasks over a 3-minute session, as described in Section~\ref{sec:n-backExperimentMethod}. The SV Task~\cite{tandon2022measuring} consisted of visual problems generated by combining four experimental factors: the number of data points, chart type, chart orientation, and task difficulty. We used two data point counts (7 and 14), four chart types (radar, bar, vertical bar, and pie), two orientations (horizontal and vertical), and three difficulty levels (easy, medium, and hard). Easy-level tasks required simple searches or direct comparisons between data values in the charts. Medium-level tasks involved identifying trends or changes, such as value increases or decreases. Hard-level tasks required participants to perform mathematical reasoning or calculations based on the presented visualizations. After each task, participants answered the NASA-TLX questionnaire survey.

\begin{figure}[t]
    \centering
    \includegraphics[width=\linewidth]{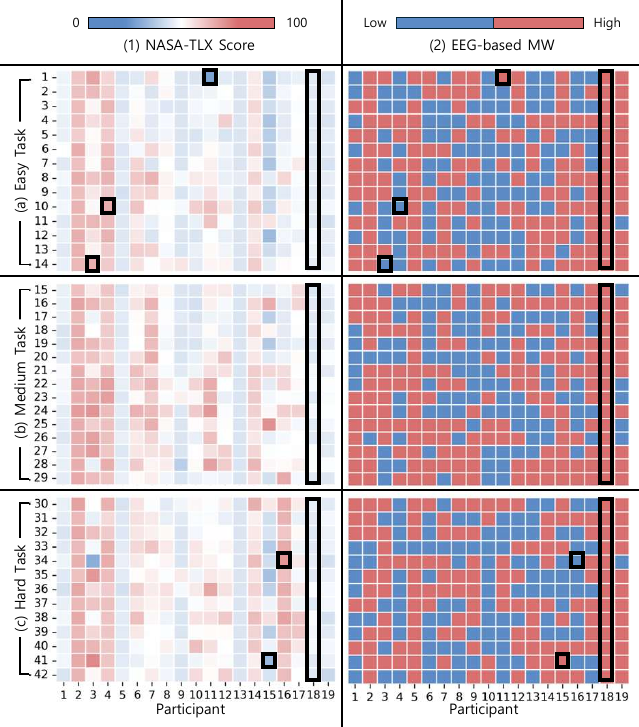}
    \caption{
        MW of participants according to task difficulty. (a), (b), and (c) are task difficulty of easy, medium, and hard levels, respectively. (1) is the NASA-TLX score, and (2) is the EEG-based MW assessment result of the $GAT_{\text{Channel\&Region}}$ model. Black boxes indicate cases with notable discrepancies between NASA-TLX and EEG-based MW estimates. Discrepancies were observed between NASA-TLX scores and EEG-based MW estimates across varying levels of task difficulty.
    }
    \label{fig:vis_compare}
\end{figure}

\subsubsection{Discrepancy Between Self-report and EEG-based MW Across Task Difficulty Levels}
\label{sec:compare_n_m}
Figure~\ref{fig:vis_compare} compares MW measured by personalized $GAT_{\text{Channel\&Region}}$ models and NASA-TLX scores across task difficulty levels defined in the SV Task. Note that the $GAT_{\text{Channel\&Region}}$ model achieved high reliability across the 19 participants, with an average accuracy of $99.38\%$, loss of $0.04$, sensitivity of $0.992$, specificity of $0.995$, and F1-score of $0.994$.
The task difficulty was categorized into (a) easy, (b) medium, and (c) hard. 
NASA-TLX in (1) and EEG-based estimations in (2) use red to indicate high MW and blue to indicate low MW.

According to the NASA-TLX results, statistically significant differences in MW were observed between conditions (a) and (b), and between (a) and (c) ($p < 0.05$), but not between (b) and (c) ($p > 0.05$). In contrast, EEG-based MW estimates revealed no significant differences across any of the task difficulty comparisons ($p > 0.05$). These findings suggest that predefined task difficulty levels do not necessarily reflect the actual cognitive effort experienced by participants. Even tasks categorized under the same difficulty level may elicit varying MW, influenced by factors such as visualization complexity, visual encoding styles, and individual differences in interpretation strategies. Overall, task difficulty, as traditionally defined, does not fully capture the range of cognitive demands involved in visualization evaluation tasks~\cite{verkennis2025predicting}.

Discrepancies between self-reported MW and EEG-based MW are observed across all levels of task difficulty in visualization tasks, as illustrated by the black boxes in Figure~\ref{fig:vis_compare}. These boxes highlight representative examples of False Negative (FN) and False Positive (FP) cases, where notable mismatches occur between NASA-TLX scores and EEG-based MW estimates. FN cases indicate high self-reported MW but low EEG-based MW, while FP cases reflect low self-reported MW but high EEG-based MW.
These cases reveal a clear mismatch in MW levels assigned by the two methods, despite being evaluated under identical task conditions.
To better understand these discrepancies, we provide quantitative evidence of inter-cluster similarity based on UMAP-derived EEG clusters in Table~\ref{tab:cluster_compare_sv}.

Table~\ref{tab:cluster_compare_sv} presents the cluster similarity between the SV task and the Spatial 1-back and 3-back tasks for the FN cases ($P_3\&T_{14}$, $P_4\&T_{10}$, $P_{16}\&T_{34}$) and FP cases ($P_{11}\&T_{1}$, $P_{15}\&T_{41}$, $P_{18}\&T_{\text{Total}}$) highlighted in Figure~\ref{fig:vis_compare}. Similarity was computed using Cohen's $d$, Mahalanobis distance, and Bhattacharyya distance based on UMAP-derived EEG clusters. $S_1$ and $S_3$ represent the similarity between the SV cluster and the 1-back and 3-back clusters, respectively. The Adjacent Cluster column indicates which of the two reference tasks the SV cluster is more similar to. EEG-based MW was estimated using the $GAT_{\text{Channel\&Region}}$ model. Table~\ref{tab:cluster_compare_sv} presents representative FN and FP cases, where (a-1)–(a-3) correspond to FN and (b-1)–(b-3) to FP. In the FN cases, SV task clusters are consistently located closer to the 1-back cluster, which is associated with lower cognitive load. Conversely, in the FP cases, SV clusters are nearer to the 3-back cluster, indicative of higher cognitive load. These spatial proximities suggest that the $GAT_{\text{Channel\&Region}}$ model aligns with cognitive effort levels observed in the $n$-back reference tasks. The consistent pattern implies that the discrepancies arise because EEG-based MW estimation captures neural activity not reflected in self-reported assessments.

\begin{table}[t]
\caption{
Cluster similarity between the SV task and the Spatial 1-back ($S_1$) and 3-back ($S_3$) tasks was computed using Cohen's $d$, Mahalanobis distance, and Bhattacharyya distance. The adjacent cluster shows which of the Spatial 1-back or 3-back EEG clusters is closest to the SV task cluster. EEG-based MW was estimated using the $GAT_{\text{Channel\&Region}}$ model.
}
\label{tab:cluster_compare_sv}
\resizebox{\columnwidth}{!}{
\begin{tabular}{|c|c|r|r|c|c|}
\hline
Participant\&Task                    & Metric        & \multicolumn{1}{c|}{$S_1$} & \multicolumn{1}{c|}{$S_3$} & Adjacent Cluster & EEG-based MW \\ \hline
\multirow{3}{*}{(a-1) $P_{3}\&T_{14}$}     & Cohen's $d$     & 2.114869     & 2.177352     & 1-back      & Low      \\ \cline{2-6} 
                                      & Mahalanobis   & 2.989527     & 3.043402     & 1-back      & Low      \\ \cline{2-6} 
                                      & Bhattacharyya & 9.172314     & 15.398117    & 1-back      & Low      \\ \hline
\multirow{3}{*}{(a-2) $P_{4}\&T_{10}$}     & Cohen's $d$    & 1.035317     & 1.612996     & 1-back      & Low      \\ \cline{2-6} 
                                      & Mahalanobis   & 2.307246     & 3.262462     & 1-back      & Low      \\ \cline{2-6} 
                                      & Bhattacharyya & 1.185847     & 4.03521      & 1-back      & Low      \\ \hline
\multirow{3}{*}{(a-3) $P_{16}\&T_{34}$}    & Cohen's $d$     & 2.301848     & 2.402963     & 1-back      & Low      \\ \cline{2-6} 
                                      & Mahalanobis   & 3.159576     & 3.538199     & 1-back      & Low      \\ \cline{2-6} 
                                      & Bhattacharyya & 7.57398      & 14.964704    & 1-back      & Low      \\ \hline
\multirow{3}{*}{(b-1) $P_{11}\&T_{1}$}     & Cohen's $d$     & 1.991462     & 0.701352     & 3-back      & High       \\ \cline{2-6} 
                                      & Mahalanobis   & 5.128987     & 3.036216     & 3-back      & High       \\ \cline{2-6} 
                                      & Bhattacharyya & 20.261181    & 2.343482     & 3-back      & High       \\ \hline
\multirow{3}{*}{(b-2) $P_{15}\&T_{41}$}    & Cohen's $d$     & 2.392093     & 2.247436     & 3-back      & High       \\ \cline{2-6} 
                                      & Mahalanobis   & 2.346177     & 2.306922     & 3-back      & High       \\ \cline{2-6} 
                                      & Bhattacharyya & 12.374136    & 7.299622     & 3-back      & High       \\ \hline
\multirow{3}{*}{(b-3) $P_{18}\&T_{\text{Total}}$} & Cohen's $d$     & 2.169244     & 1.936149     & 3-back      & High       \\ \cline{2-6} 
                                      & Mahalanobis   & 2.437846     & 2.263851     & 3-back      & High       \\ \cline{2-6} 
                                      & Bhattacharyya & 15.540598    & 4.976519     & 3-back      & High       \\ \hline
\end{tabular}}
\end{table}

\begin{figure}[t]
    \centering
    \includegraphics[width=0.95\columnwidth]{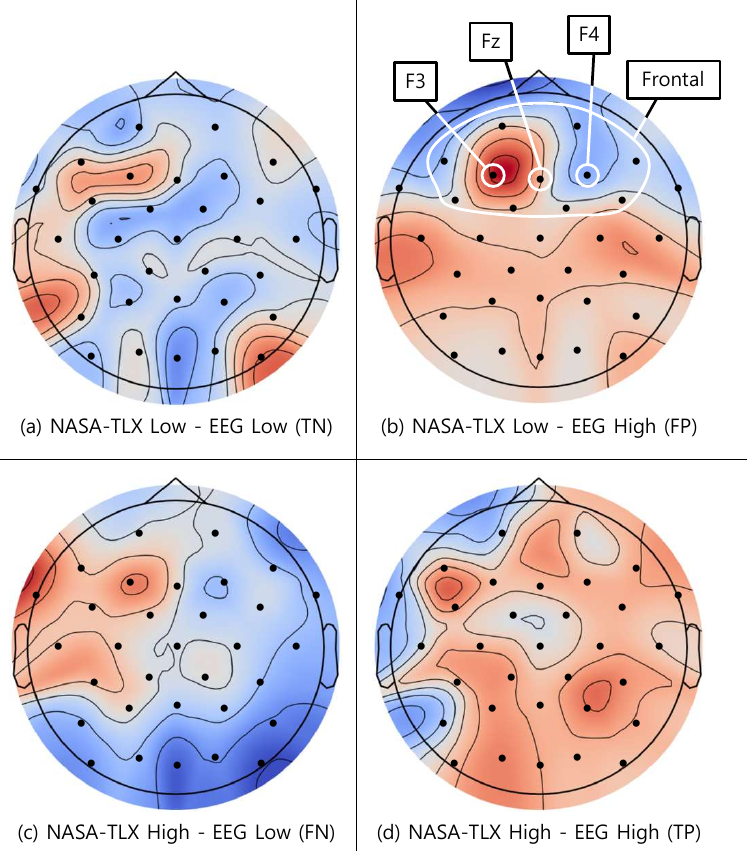}
    \caption{
        Topographic maps of average theta activity. Black dots indicate EEG channel locations.
        Topographic maps illustrate cases categorized by the agreement between self-reported and EEG-based MW: (a) true negative (TN), (b) false positive (FP), (c) false negative (FN), and (d) true positive (TP). The averaged theta activity across these four conditions suggests the presence of unconscious cognitive effort not captured by self-reported MW.
    }
    \label{fig:eeg_topo}
\end{figure}

\subsubsection{Discrepancies due to Unconscious Cognitive Effort}
\label{sec:uce}
Another significant factor underlying discrepancies between EEG-based and self-reported MW measures involves unconscious cognitive effort, defined as cognitive control processes that operate beyond participants' conscious awareness while still impacting neural responses. To investigate this phenomenon, we examined average theta activity topographies across four MW classification categories, as presented in Figure~\ref{fig:eeg_topo}: (a) True Negative (TN), (b) False Positive (FP), (c) False Negative (FN), and (d) True Positive (TP).

Each topographic map illustrates spatial distributions of theta-band (4–7 Hz) activity. A consistent color scale is used across all maps, with red indicating higher theta activity and blue indicating lower activity. The overlaid black dots mark EEG electrode locations based on the Emotiv Epoc Flex configuration shown in Figure~\ref{fig:sensorPlacement}. Notably, the frontal-midline region—including channels F3, Fz, and F4—is widely associated with cognitive control and mental effort~\cite{mcferren2021causal, smith2001monitoring}. Across all conditions, the F3 channel exhibits consistent baseline activation, likely reflecting general task engagement. However, differences in activation intensity and spatial extent across the four conditions reveal distinct neural signatures that help characterize the nature of the discrepancies between EEG-based and self-reported MW measures.

Figure~\ref{fig:eeg_topo} (d), corresponding to the TP condition where both NASA-TLX and EEG-based measures indicate high MW, exhibits the most pronounced and widespread frontal-midline theta activity. This activation pattern is consistent with prior studies linking elevated frontal theta power to heightened cognitive demand~\cite{aksu2024mental}. In contrast, the TN condition in Figure~\ref{fig:eeg_topo} (a), where both measures indicate low workload, shows minimal and diffuse theta activity, reflecting low cognitive engagement. These results demonstrate that congruent MW assessments are associated with distinct and interpretable neural activation patterns.

In Figure~\ref{fig:eeg_topo} (b), representing the FP condition, a notable discrepancy is observed. Although participants self-reported low MW, frontal theta activity remains relatively elevated, particularly over the F3 and Fz channels.
This pattern indicates cognitive engagement that was not consciously perceived by the participants.
The similarity of this activation pattern to that of the TP condition suggests underlying cognitive processes that contribute to MW but are not reflected in subjective reports. These findings highlight the limitations of self-reported measures and underscore that certain aspects of cognitive effort operate below the threshold of conscious awareness.

In contrast, Figure~\ref{fig:eeg_topo} (c), depicting the FN condition, reveals low frontal theta activation despite high self-reported mental workload. Compared to the FP case in (b), theta activity at F3 and Fz is noticeably reduced, indicating that the subjective perception of effort was not paralleled by corresponding neural engagement. This divergence in topographic patterns between (b) and (c), despite both representing discrepancies between EEG-based and self-reported measures, suggests that only a subset of such mismatches can be attributed to unconscious cognitive effort, as reflected by elevated frontal theta in (b).

These findings are consistent with theoretical frameworks that conceptualize mental workload as comprising both conscious and unconscious components~\cite{mondellini2024multimodal}. 
Frontal-midline theta activity has been linked to internal cognitive control and goal-directed effort~\cite{mcferren2021causal}.
The elevated frontal theta observed in FP cases highlights the contribution of unconscious cognitive effort to task engagement, even when such effort is not accessible to self-report or introspection. Accounting for these implicit cognitive processes is critical for resolving discrepancies between different MW measures and for achieving a more comprehensive understanding of MW in complex domains such as visual analytics.

%% file: 6_discussion.ins
\section{Discussion}
\label{sec:discussion}
In studies examining discrepancies between self-reported and EEG-based MW assessments, the architecture of the $GAT_{\text{Channel\&Region}}$ model plays a crucial role in ensuring both experimental validity and interpretability. This study considers two key architectural factors: (1) the selection of appropriate EEG features and (2) the choice between a personalized model tailored to individual participants and a participant-independent model designed to generalize across participants. These decisions represent ongoing challenges in EEG research, as the optimal configuration often depends on the specific objectives and contextual constraints of a given study. The following sections present the rationale underlying the model design choices adopted in this work.

\begin{figure}[t]
    \centering
    \includegraphics[width=\linewidth]{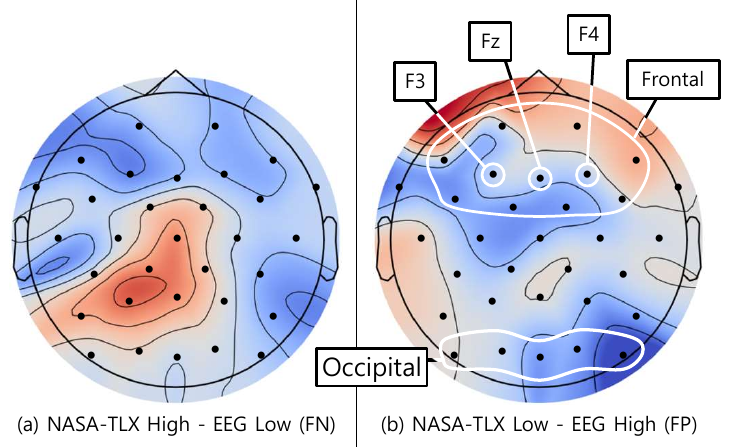}
    \caption{
    Topographic maps of average alpha activity.
    Black dots indicate EEG channel locations. 
    (a) and (b) represent discrepancy cases between self-reported and EEG-based MW: (a) corresponds to a false negative (FN) and (b) to a false positive (FP). The observed alpha band deactivation in both cases suggests that MW may be effectively measured using the alpha band.
    }
    \label{fig:eeg_topo_t_a}
\end{figure}

\subsection{Feature Selection from EEG Signals}
Section~\ref{sec:compare_n_m} analyzed the discrepancies between MW estimates derived from the $GAT_{Channel\&Region}$ model and those obtained via the NASA-TLX. The $GAT_{Channel\&Region}$ model was trained using theta band signals as input features. We selected theta band signals based on prior studies that have consistently demonstrated its strong association with cognitive workload~\cite{smith2001monitoring, aksu2024mental, zahabi2025cognitive}. However, not only theta band but also the alpha band has shown robust performance in estimating MW~\cite{klimesch1996memory, klimesch1999eeg}. In this section, we discuss how the choice of EEG features influences the interpretation of discrepancy experiments in MW estimation.

Figure~\ref{fig:eeg_topo_t_a} illustrates topographic maps of average alpha activity, highlighting discrepancy cases such as the FN case in (a) and the FP case in (b). A consistent color scale is applied across all maps, with red indicating higher alpha activity and blue indicating lower activity.
In general, theta and alpha bands exhibit inverse activation patterns about MW~\cite{raufi2022evaluation, yin2017cross}. Specifically, theta  activity in the frontal region (e.g., F3, Fz, F4) tends to increase under higher cognitive control demands, whereas alpha activity decreases. In the occipital region, alpha activity is also typically suppressed under high cognitive demands~\cite{klimesch1999eeg}.

Considering regional activation patterns of the alpha band, a comparison between Figures~\ref{fig:eeg_topo_t_a} (a) and (b) reveals an inverse pattern in the frontal region: increased alpha activity suppression complements the elevated theta activity discussed in Section~\ref{sec:uce}, suggesting heightened cognitive effort not fully captured by participants’ subjective reports. In the occipital region of Figure~\ref{fig:eeg_topo_t_a}(b), the observed alpha deactivation aligns with theoretical expectations, further reinforcing the findings presented in Section~\ref{sec:uce}. However, unlike theta activity, which is widely recognized as a robust neurophysiological marker of MW, alpha activity responses are more sensitive to various mental states, including task disengagement, fatigue, and rest~\cite{li2024research, raufi2022evaluation}. This sensitivity introduces some uncertainty in using alpha activity as a reliable MW indicator.

As shown in Figure~\ref{fig:eeg_topo_t_a}, the experimental interpretation based on the alpha activity produced results comparable to those derived from the theta activity, suggesting that theta is not the sole viable option for EEG-based MW research. This interpretive similarity likely stems from the indirect relationship between alpha activity and MWs.
However, alpha activity is susceptible to broader contextual and physiological influences, such as disengagement, fatigue, and resting states, which can complicate controlled and consistent interpretation in experimental settings. 
For these reasons, we selected theta band as the primary input feature for MW estimation and interpretation in our visualization tasks.

Nonetheless, the alpha band provides complementary information by capturing task-specific brain activities such as visual stimulus processing and attentional shifts~\cite{sauseng2005shift}. 
Therefore, we plan to incorporate these diverse EEG features in future work to conduct more detailed analyses of discrepancies arising from different MW measurement methods.
Moreover, other EEG-based features, such as the $\frac{\text{Theta}}{\text{Alpha}}$ ratio~\cite{aksu2024mental, anders2024unobtrusive}, emphasize the limitations of relying on a single frequency band. Incorporating a broader range of EEG features may provide more comprehensive insights into MW during experimental design and analysis. Consequently, future work will integrate these diverse features to enable more detailed analyses of discrepancies among different MW measurement methods.

\subsection{Model Selection Across Levels of Personalization}
\begin{figure}[t]
    \centering
    \includegraphics[width=\linewidth]{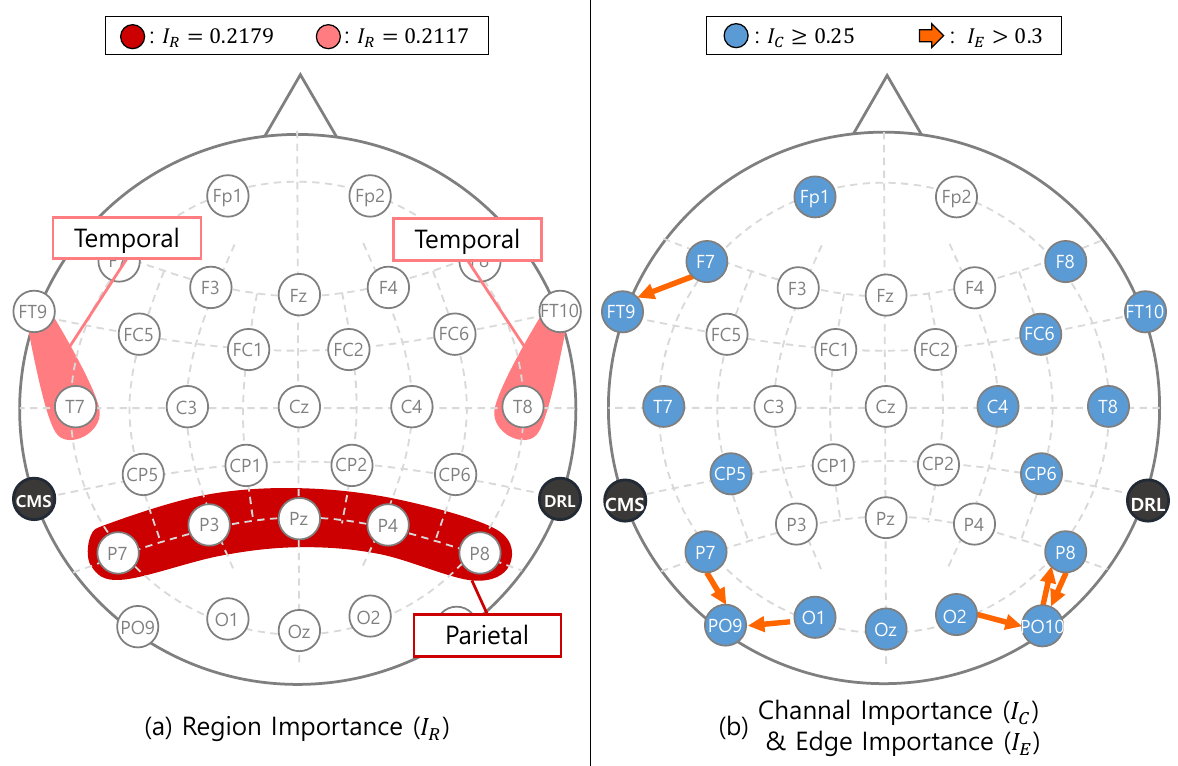}
    \caption{
        Attention-based interpretability analysis of the $GAT_{\text{Channel\&Region}}$ model for MW estimation. (a) presents region importance derived from the region graph learning block, while (b) illustrates channel and edge importance obtained from the adjacent channel graph learning block. 
        Common attention features across all personalized models indicate consistent EEG responses associated with MW.
    }
    \label{fig:brain_activity}
\end{figure}

\begin{table}[t]
\caption{
    MW estimation performance of the participant-independent model using leave-one-subject-out (LOSO) cross-validation. SD indicates standard deviation.
}
\label{tab:general_model}
\resizebox{\columnwidth}{!}{
\begin{tabular}{|c|r|r|r|r|r|}
\hline
Metric      & \multicolumn{1}{c|}{Mean}    & \multicolumn{1}{c|}{SD}     & \multicolumn{1}{c|}{Median} & \multicolumn{1}{c|}{Min} & \multicolumn{1}{c|}{Max}     \\ \hline
Accuracy (\%)    & 50.62 & 27.89 & 50   & 0 & 99.94 \\ \hline
Loss        & 1.534    & 2.039   & 0.762   & 0.003   & 8.279    \\ \hline
Sensitivity & 0.581    & 0.458    & 0.925   & 0   & 1       \\ \hline
Specificity & 0.435    & 0.444    & 0.297    & 0   & 1       \\ \hline
F1 Score    & 0.476    & 0.359    & 0.632   & 0   & 0.999       \\ \hline
\end{tabular}}
\end{table}

In this study, we developed a personalized model based on $GAT_{\text{Channel\&Region}}$ to estimate MW from EEG data collected on a per-participant basis. This personalized modeling approach allows for a detailed analysis of discrepancies between self-reported and EEG-based MW estimates. From a neuroergonomics perspective, EEG signals inherently capture individual neurophysiological characteristics~\cite{baldwin2012adaptive, milos2024towards}, supporting the need for personalized models in cognitive state estimation.

Prior to adopting the personalized model, we evaluated a participant-independent model using leave-one-subject-out (LOSO) cross-validation~\cite{kingphai2024mental}. In this setup, the participant-independent model was trained on EEG data from 18 participants and tested on the remaining participant, with the process repeated for all 19 participants. While LOSO provides a robust framework for subject-independent evaluation, the participant-independent model exhibited unstable performance on the visualization task, achieving a mean accuracy of 50.62\%, a median of 50.00\%, and a large standard deviation of 27.89\%. Accuracy varied significantly across individuals, ranging from 0.00\% to 99.94\%, reflecting substantial inter-subject variability. Although participant-independent models offer scalability and ease of deployment to new users and tasks~\cite{yedukondalu2025cognitive}, their inconsistent performance in our evaluation motivated the use of personalized modeling. This shift enabled a more stable and participant-sensitive estimation of MW, better accounting for individual neurophysiological differences.

Despite adopting personalized modeling in our experiments, attention-based analysis revealed notable consistencies in the prioritization of EEG features across participants. These recurring patterns suggest the presence of shared neural mechanisms underlying MW, as reflected by common EEG signal characteristics. Such consistencies open the possibility of generalizing these insights to inform the design of more robust participant-independent models. 
To investigate this potential, we conducted an attention-based feature importance analysis, examining region importance ($I_R$), channel importance ($I_C$), and edge importance ($I_E$) derived from the $GAT_{\text{Channel\&Region}}$ model, as illustrated in Figure~\ref{fig:brain_activity}. This visualizes the most salient components identified by the model. (a) shows regions with prominent $I_R$ values computed from the region graph learning block. (b) shows channels and edges with prominent $I_C$ and $I_E$ values derived from the channel and edge attention mechanisms in the adjacent channel graph learning block.

For region importance ($I_R$), as shown in (a), calculated as the average attention score of edges connected to each region node in the region-level graph, the parietal and temporal regions consistently exhibited the highest values across participants ($I_R = 0.2177$ and $0.2117$, respectively). This pattern suggests that the model heavily relied on signals from the parietal cortex, associated with higher-order cognitive processing, and the temporal cortex, known for its role in multisensory integration~\cite{souza2023neuropsychology, hirst2022multisensory}. These findings align with the established neurocognitive functions of these regions and highlight their centrality in MW estimation.

For channel importance ($I_C$), as indicated by the blue circles in (b), defined as the normalized attention scores of edges connected to each EEG channel node, consistently high values were observed in parietal channels (PO9, PO10, O1, O2) and temporal channels (FT9, T7, FT10, T8), all with $I_C \geq 0.2$. This pattern further supports the prominent role of parietal and temporal regions in MW estimation. The elevated attention assigned to these channels suggests that EEG signals from these areas are particularly informative, likely due to their involvement in higher-order cognitive processing and multisensory integration, respectively.

Edge importance ($I_E$), highlighted by the orange arrows in (b) and calculated from the attention scores assigned to connections between EEG channels, revealed consistent patterns across participants. 
Notably, high importance values ($I_E \geq 0.3$) were observed in edges linking the frontal–temporal region (F7–FT9), parietal–occipital region (P7–PO9, P8–PO10), and occipital region (O1–PO9, O2–PO10).
This pattern signifies that the cognitive and visual information processing demands of the visualization task commonly activate interactions between the frontal cognitive control functions and the temporal sensory integration functions, as well as between the occipital higher-order cognitive functions and the parietal visual information processing capabilities~\cite{formoso2021detecting, diaz2020eeg, souza2023neuropsychology}.

Overall, these findings suggest that, despite inter-individual variability, consistent spatial patterns represent the neural encoding of mind wandering across brain regions and EEG channels. Such shared attentional features lay a foundation for advancing participant-independent MW estimation models in the future.

%% file: 7_conclusion.ins
\section{Conclusion}
This paper examined discrepancies between EEG-based and self-reported mental workload (MW) assessments within the context of data visualization. 
Preliminary analyses revealed inconsistencies between MW estimates derived from theta power in EEG signals and self-reported NASA-TLX scores, alongside weak correlations with behavioral measures.
Motivated by these initial findings, two follow-up experiments were conducted employing a personalized EEG model based on Graph Attention Networks (GAT). This model was applied to the Visualization Literacy Assessment Test (VLAT)~\cite{lee2016vlat} and the Spatial Visualization (SV) Task~\cite{tandon2022measuring}, leveraging graph-based representations to learn tailored regional and channel-level EEG features. 
The experiments investigated MW discrepancies in relation to EEG-based MW and task difficulty, divergences between EEG-based and self-reported MW across difficulty levels, and the influence of unconscious cognitive effort not captured by self-reports.
As discussed in Section~\ref{sec:discussion}, future work will incorporate multiple frequency bands to enable a more nuanced characterization of MW and leverage consistent attention-based patterns to develop a generalized, participant-independent MW estimation model applicable across diverse tasks and populations.

\section{Ethics Statement}
This study was approved by the Institutional Review Board (IRB) of Sejong University, Seoul, South Korea (Approval Number: SUIRB-HR-2024-011; Approval Date: August 27, 2024). Prior to participation, all subjects were fully informed about the purpose, procedures, potential risks, and benefits of the study. Written informed consent was obtained from all participants, confirming their voluntary participation. EEG data were collected in accordance with the international 10–20 system, and all personally identifiable information was anonymized to protect participant privacy and ensure data confidentiality. The collected EEG data were securely encrypted and used solely for research purposes. All procedures in this study were conducted in accordance with the ethical principles outlined in the Declaration of Helsinki.